\documentclass[twocolumn]{aastex631}

\usepackage{amsmath}

\usepackage{xcolor,colortbl}

\begin{document}


\title{Characterization of the Trans-Alfv\'enic Region Using Observations from Parker Solar Probe}

\author[0000-0002-0786-7307]{Subash Adhikari*}
\affiliation{Department of Physics and Astronomy, University of Delaware, Newark, DE 19716 USA}
\email{*subash@udel.edu}

\author[0000-0002-6962-0959]{Riddhi Bandyopadhyay}
\affiliation{Department of Physics and Astronomy, University of Delaware, Newark, DE 19716 USA}

\author[0000-0002-5354-1164]{Joshua Goodwill}
\affiliation{Department of Physics and Astronomy, University of Delaware, Newark, DE 19716 USA}

\author[0000-0001-7224-6024]{William H. Matthaeus}
\affiliation{Department of Physics and Astronomy, University of Delaware, Newark, DE 19716 USA}
\author[0000-0003-3414-9666]{David Ruffolo}
\affiliation{Department of Physics, Faculty of Science, Mahidol University, Bangkok 10400, Thailand}

\author[0000-0001-9597-1448]{Panisara Thepthong}
\affiliation{LPC2E, OSUC, CNRS, University of Orléans, CNES, Orléans F-45071, France}

\author[0000-0002-6609-1422]{Peera Pongkitiwanichakul}
\affiliation{Department of Physics, Faculty of Science, Kasetsart University, Bangkok 10900, Thailand}

\author[0000-0003-3891-5495]{Sohom Roy}
\affiliation{Department of Physics and Astronomy, University of Delaware, Newark, DE 19716 USA}

\author[0000-0003-4168-590X]{Francesco Pecora}
\affiliation{Department of Physics and Astronomy, University of Delaware, Newark, DE 19716 USA}

\author[0000-0002-7174-6948]{Rohit Chhiber}
\affiliation{Heliophysics Science Division, NASA Goddard Space Flight Center, Greenbelt, MD 20771, USA}
\affiliation{Department of Physics and Astronomy, University of Delaware, Newark, DE 19716 USA}

\author[0009-0005-9366-6163]{Rayta Pradata}
\affiliation{Department of Physics and Astronomy, University of Delaware, Newark, DE 19716 USA}

\author[0000-0002-0209-152X]{Arcadi Usmanov}
\affiliation{Heliophysics Science Division, NASA Goddard Space Flight Center, Greenbelt, MD 20771, USA}
\affiliation{Department of Physics and Astronomy, University of Delaware, Newark, DE 19716 USA}

\author[0000-0002-7728-0085]{Michael Stevens}
\affiliation{Center for Astrophysics, Harvard \& Smithsonian, Cambridge, MA 02138, USA}

\author[0000-0002-6145-436X]{Samuel Badman}
\affiliation{Center for Astrophysics, Harvard \& Smithsonian, Cambridge, MA 02138, USA}

\author[0000-0002-4559-2199]{Orlando Romeo}
\affiliation{Department of Earth \& Planetary Science and Space Sciences Laboratory, University of California at Berkeley, Berkeley, CA 94720, USA}

\author[0009-0008-8723-610X]{Jiaming Wang}
\affiliation{Department of Physics and Astronomy, University of Delaware, Newark, DE 19716 USA}

\author[0000-0002-5317-988X]{Melvyn L. Goldstein}
\affiliation{Space Science Institute, Boulder, CO 80301, USA}







\begin{abstract}
Close to Earth the solar wind is usually super-Alfv\'enic, i.e. the speed of the solar wind is much larger than the Alfv\'en speed. 
However, in the lower coronal regions, the solar wind is mostly sub-Alfv\'enic.
With the Parker Solar Probe (PSP) crossing the boundary between the sub- and super-Alfv\'enic flow,~\citet{bandyopadhyay2022sub} performed a turbulence characterization of the sub-Alfv\'enic solar wind with initial data from encounters $8$ and $9$. In this study, we re-examine the turbulence properties such as turbulence amplitude, anisotropy of the magnetic field variance, intermittency and switchback strength extending with PSP data for encounters $8$--$19$. The later orbits probe lower altitudes and experience sub-Alfv\'enic conditions more frequently providing a greater statistical coverage to contrast sub- and super-Alfv\'enic solar wind. Also, by isolating the intervals where the solar wind speed is approximately equal to the Alfv\'en speed,
we explore the transition in more detail. We show that the amplitude of the normalized magnetic field fluctuation is smaller for the sub-Alfv\'enic samples. While solar wind turbulence in general is shown to be anisotropic, the sub-Alfv\'enic samples are more anisotropic than the super-Alfv\'enic samples, in general.
Further, we show that the sub- and super-Alfv\'enic samples do not show much distinction in terms of intermittency strength. Finally, 
consistent with prior results, we find no evidence for polarity reversing 
$> 90^\circ$ switchbacks in the sub-Alfv\'enic solar wind.

\end{abstract}



\section{Introduction} \label{sec:intro}

Parker Solar Probe (PSP)~\citep{fox2016solar,raouafi2023parker}
explores the solar atmosphere more closely than any previous {\it in situ} mission and is therefore able to observe
processes potentially 
responsible for 
heating and accelerating the 
solar wind plasma.
This already iconic dataset finally enables probing of the origin of the solar wind-- a problem that has challenged the community for more than half a century. 
It is now well understood that the plasma close to the sun is magnetically controlled and at least approximately corotates with the sun. 
At large distances the solar magnetic field can no longer control the plasma as the energy density in the flow greatly exceeds the thermal and magnetic energy densities. 
Between these limits are several interesting transitions where first, the flow energy density exceeds the magnetic energy density, and where second, the thermal pressure equals or exceeds the magnetic pressure. The first of these is commonly called the
Alfv\'en critical point \citep{weber1967angular}, now thought to be a more complex region with numerous such transitions \citep{chhiber2024alfven}. 
Exploring this ``Alfv\'en 
transition region" in PSP data provides a unique opportunity to understand the differences  in physical properties and processes that operate in the corona (the region below the transition) and the extended solar wind above.

In regard to exploring the change in conditions across the trans-Alf\'ven transition, 
\citet{bandyopadhyay2022sub}
contrasted the properties of sub-Alfv\'enic solar wind as measured by PSP during the eighth (E8) and ninth (E9) solar encounters. 
The sub-Alfv\'enic datasets sampled in these earlier encounters provided a first 
opportunity to examine the sub-Alfv\'enic properties of the upper corona or
sub-Alfv\'enic solar wind.

Here we extend that study to datasets from encounters E8 to E19.
These later orbits probe lower altitudes and experience sub-Alfv\'enic conditions more frequently.
With greater statistical coverage, we are able to better contrast the sub- and super-Alfv\'enic solar wind conditions.
Furthermore,
by isolating the intervals where the solar wind is trans-Alfv\'enic, (i.e. Alfv\'en Mach number $M_A\simeq 1$), we explore the transition in more detail. 
Specifically, 
we analyze 
solar wind in categories 
defined as 
sub-Alfv\'enic
$M_A < 0.85$, trans-Alfv\'enic
$0.85 \leq M_A \leq 1.15$, and super-Alfv\'enic 
$M_A > 1.15$ regimes and study their properties. Note that using a narrower range for the trans-Alfv\'enic regime, such as $0.9 \leq M_A \leq 1.1$, yields qualitatively similar results.

We note that other studies have also explored the trans-Alfv\'en transition 
\citep{zank2022turbulence,zhao2022turbulent,jiao2023properties}
adopting strategies 
that are complementary to ours. 
The paper is organized 
along the lines of~\citet{bandyopadhyay2022sub}:
Section~\ref{sec:data} describes the datasets employed. 
Section ~\ref{sec:results} provides details of the results on turbulence amplitudes
\citep{bruno2013solar}, 
variance anisotropy \citep{oughton2016variance},
intermittency measured using partial variance of increment (PVI) methods \citep{greco2018partial},
and occurrence rate 
of the switchback parameter~\citep{de2020switchbacks}. Finally, in Section \ref{sec:discussion} we present our conclusions and discuss the implications.

\section{Data During Encounters $8-19$} \label{sec:data}
Magnetic-field ($\mathbf{B}$) data in the PSP measurements are obtained from the flux-gate magnetometer on the FIELDS instrument suite~\citep{bale2016fields,bale2019highly,bale2020psp}. 
The proton radial velocities $(V_R)$ are obtained from the Solar Probe ANalyzer for Ions
(SPAN-I) on the SWEAP instrument suite~\citep{kasper2016solar,kasper2019alfvenic,livi2020psp}. The FIELDS QTN electron densities
$(N_e)$ are computed from the quasi-thermal noise (QTN) spectrum~\citep{romeo2023near}
measured by the Radio Frequency Spectrometer onboard PSP~\citep{moncuquet2020first}. With these measurements, the local Alfvén speed is calculated
as $V_A = \lvert \mathbf{B}\rvert/\sqrt{\mu_0 m_p N_e}$ , where $\mu_0$ is the magnetic permeability of a vacuum and $m_p$ is the proton mass. Finally, the local Alfv\'en Mach number is estimated as $M_A = V_R/V_A$ with a resolution of $60~\rm{s}$. The data is then divided into sub-Alfv\'enic, trans-Alfv\'enic and super-Alfv\'enic {\it regions} as intervals in which the sampled data satisfy the criteria $M_A<1$, $0.85\leq M_A \leq 1.15$, and $M_A>1$ respectively, each lasting for a duration of $10$ minutes. Further, all the calculations involving magnetic field uses a dataset of time resolution of $0.25$ \rm{s}, which are later resampled to map with the Alfv\'en Mach number $M_A$.  

\begin{figure}
    \centering    \includegraphics[width=1.05\linewidth]{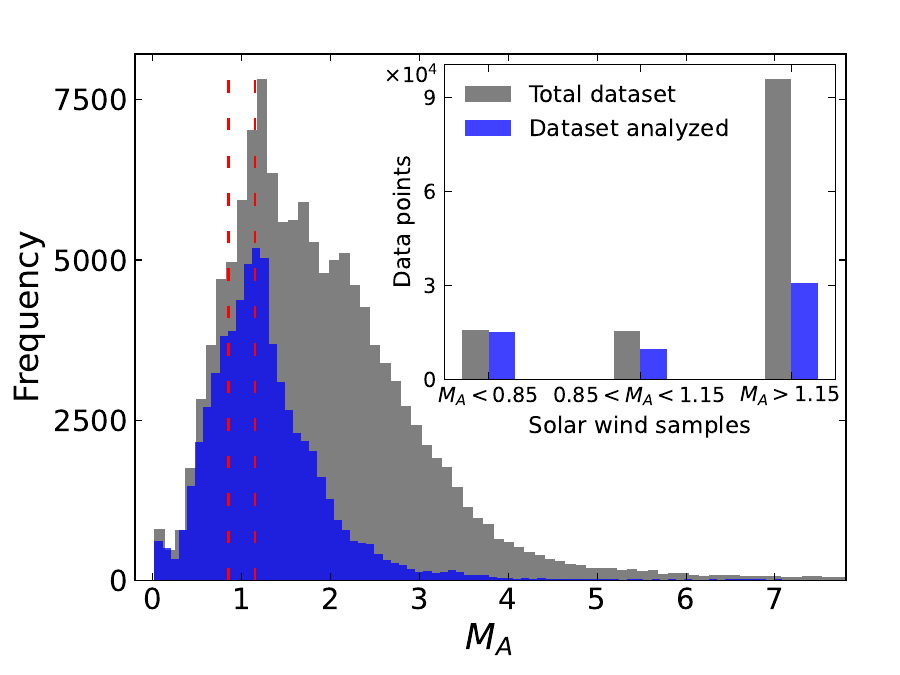}
    \caption{Comparison of distributions of the Mach number $M_A$ for the analyzed dataset shown in Fig.~\ref{fig:PSP_data} with the total datasets in encounters $8-19$. The data set is divided into sub-Alfv\'enic ($M_A<0.85$), trans-Alfv\'enic ($0.85\leq M_A\leq 1.15$) and super-Alfv\'enic ($M_A>1.15$) samples by the red vertical dashed lines. Embedded: Comparision of the number of datasets in each category (sub-Alfv\'enic, trans-Alfv\'enic and super-Alfv\'enic).}
    \label{fig:mach_pdf}
\end{figure}

A graphical overview of the datasets collected in each encounter is given in the Appendix with the start and end times for each encounter listed in Table~\ref{tab:tableI} (See Appendix). In Fig.~\ref{fig:mach_pdf}, we compare the total datasets available in these encounters with the subset chosen for the analysis. For the purpose of this study, we use a total of 4 days of data for each encounter (enc) around the perihelion (two days on either side) to create a comparable sample size for both sub- and super-Alfv\'enic datasets and avoid statistical bias. Note that almost all of the sub-Alfv\'enic periods sampled by the PSP during these encounters have been included as shown in the bar chart embedded in Fig.~\ref{fig:mach_pdf}. 
Many super-Alfv\'enic periods distant 
from the sun are excluded from our analysis. 


\section{Results} \label{sec:results}
\subsection{Turbulence Amplitude}
\begin{figure*}
    \centering
    \includegraphics[width=1\linewidth]{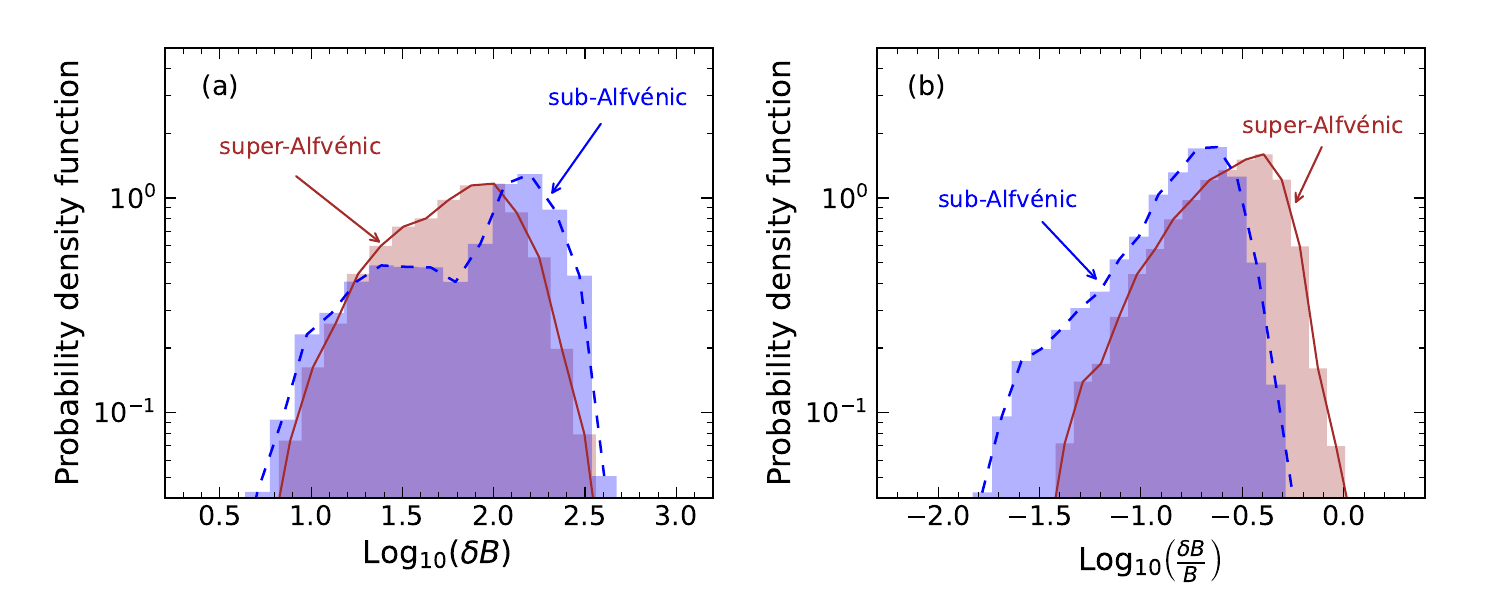}
    \caption{Probability distribution function (PDF) of  (a) the magnetic-field turbulence amplitude (in nT) and (b) the normalized magnetic-field turbulence amplitude in the sub-Alfvénic and super-Alfvénic solar wind intervals as observed by the PSP in encounters $8-19$.}
    \label{fig:dB_byB_pdf}
\end{figure*}

First, we examine the amplitude of the magnetic field fluctuations. 
In Fig.~\ref{fig:dB_byB_pdf}a, we show the probability distribution function (PDF) of the turbulence amplitude, calculated as
$\delta B= \sqrt{\langle\lvert \mathbf{B}(t)-\langle \mathbf{B} \rangle \rvert^2 \rangle}$,
where $\langle \cdots \rangle$ is a time average, over some chosen time range, usually several correlation times.  Throughout this paper, we choose averaging
intervals of $10$ minutes (on the order of few correlation times; see 
\citet{parashar2020measures}) 
and evaluate the turbulence amplitude in each interval. Then, we separately accumulate the intervals classified as 
sub-Alfv\'enic, trans-Alfv\'enic or super-Alfv\'enic.
In Fig \ref{fig:dB_byB_pdf}a we show the frequency of occurrence of values of $\delta B$ (in $nT$),
for super-Alfv\'enic and sub-Alfv\'enic intervals. We exclude the PDF of $\delta B$ near unit Alfv\'en Mach number to clearly distinguish the two population.
This result differs from that of~\citet{zank2022turbulence,bandyopadhyay2022sub}. In particular we cannot conclude that the sub-Alfv\'enic period have systematically lower $\delta B$ as was reported in that earlier study. In fact there is a suggestion of a second population of sub-Alfv\'enic contributions with corresponding $\text{Log}_{10}(\delta B) \sim 2.2$. Inspection of the results from each encounter (not shown) indicates that this feature is mainly due to several encounters, such as E12, E17, E18, and E19. 
Most of the other encounters are more consistent with the conclusions in~\citet{bandyopadhyay2022sub}. 
Fluctuations in the radial magnetic field ($B_r$) are also small for sub-Alfv\'enic intervals and follow~\citet{bandyopadhyay2022sub}. But the 
fluctuations in the normal ($B_n$) and tangential ($B_t$) magnetic fields are larger in the sub-Alfv\'enic intervals. 
Since the contribution to this secondary peak are mainly due to later encounters, we may 
hypothesize that larger amplitude transverse fluctuations are to be found in 
regions closer to the sun and perhaps inside the region of complex trans-Alfv\'enic 
transitions~\citep{bandyopadhyay2022sub}. 
Because of this potential complication, 
we examine now the turbulence amplitude parameter $\delta B /B$, where $B$ is the average of the magnitude of the magnetic field over the interval i.e. $B=\langle \lvert B \rvert \rangle_{\text{interval}} $.

In Fig.~\ref{fig:dB_byB_pdf}b we show the frequency of occurrence of the normalized turbulence level (in logarithm), for the sub- and super-Alfv\'enic intervals. Clearly, the normalized turbulence amplitudes for the super-Alfv\'enic samples are larger relative to 
the sub-Alfv\'enic intervals. 
The most probable values of $\delta B/B $ for the sub-Alfv\'enic, and super-Alfv\'enic wind are
$0.20$ and $0.38$, respectively indicating stronger turbulence in super-Alfv\'enic samples compared with sub-Alfv\'enic samples. Similarly, the average value of the normalized turbulence amplitude for the sub-Alfv\'enic solar wind is $0.18$ with a standard deviation $\sigma=0.10$. For the super-Alfv\'enic case, the average value of the normalized turbulence amplitude is $0.30$ with $\sigma=0.17$. While the trans-Alfv\'enic datasets are not included in these plots, the average and most probable values lie between the two regimes and are listed in Table~\ref{table:average_values} for comparison.

\subsection{Variance Anisotropy}
In this subsection, we examine the variance anisotropy~\citep{oughton2016variance,parashar2016variance} of the magnetic field.
Variance anisotropy measures the departure from equipartition of 
the magnetic field fluctuation energies in each cartesian 
component. For variance isotropy, 
$\langle b_x^2 \rangle=\langle b_y^2 \rangle=\langle b_z^2 \rangle$. 
Here, the mean magnetic field is excluded, usually by subtracting the average value $\mathbf{B}_0$. In a cartesian coordinate system, a measure 
of variance anisotropy is often defined as
\begin{equation}
    A_b = \frac{\langle b_x^2+b_y^2\rangle}{\langle b_\parallel^2\rangle},
\end{equation}
where $\mathbf{B}_0$ is chosen to be along the z-axis and the magnetic fluctuations are $\textbf{b} = (b_x, b_y, b_\parallel)$. For an isotropic distribution of magnetic field components, the variance anisotropy takes on a value of $A_b = 2$.

\begin{figure}
    \centering
    \includegraphics[width=1\linewidth]{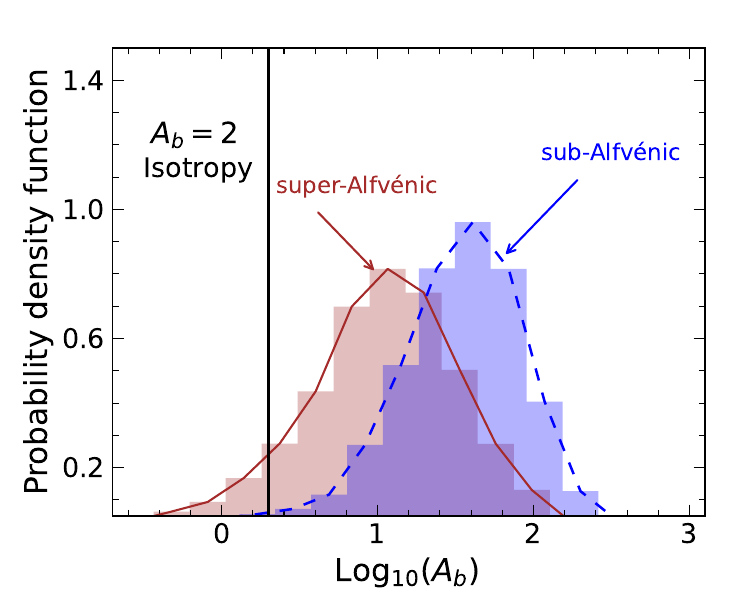}
    \caption{Probability distribution function of variance anisotropy in the sub-Alfv\'enic and super-Alfv\'enic solar wind intervals observed by PSP in encounters $8-19$. The vertical solid (black) line indicates the value $A_b = 2$, which corresponds to isotropic distribution.}
    \label{fig:Anisotropy_pdf}
\end{figure}

The variance anisotropy $A_b$ is calculated in each interval of $10$-minute duration
for the PSP datasets described in Section~\ref{sec:data}. The histograms of these magnetic field anisotropies for the sub-Alfvénic and super-Alfvénic periods are shown in Fig.~\ref{fig:Anisotropy_pdf}. Again the intermediate trans-Alfv\'enic intervals are excluded from the histograms for clarity; with the average and most probable
values shown in Table (\ref{table:average_values}).

Clearly, all these three selected regimes of the solar wind are highly anisotropic
compared to isotropy, for which one would expect  
$A_b \sim 2$, and $\text{Log}_{10}(A_b)\simeq 0.3$. 
However, the sub-Alfv\'enic solar wind periods display the strongest variance
anisotropy of the three groups. 
The super-Alfv\'enic samples show the least variance anisotropy. 
From these distributions, one can also find the most probable value of the variance anisotropy $\Tilde{A}_b=56.23$ for the sub-Alfv\'enic case, and $\Tilde{A}_b=13.18$ for the super-Alfv\'enic case. Similarly, the average values of the variance anisotropy for the sub-Alfv\'enic case is $\overline{A}_b=51.09$ with a standard deviation of $\sigma_{A_b}=65.50$. For the super-Alfv\'enic samples, $\overline{A}_b=20.16$ with $\sigma_{A_b}=30.33$.
The degree of variance anisotropy is clearly a distinguishing feature of
the sub-Alfv\'enic corona as explored by PSP.

\subsection{Intermittency}

Intermittent signatures can be partially quantified by the partial variance of increment (PVI) method, a measure that detects the occurrence of sharp gradients in quantities such as the magnetic field~\citep{greco2008intermittent,greco2018partial}. 
The PVI of the magnetic field $\mathbf{B}$ for a time lag of $\tau$ at a time $t$ is defined as
\begin{equation}
    \text{PVI}_{t,\tau}=\frac{\lvert\Delta \mathbf{B}(t, \tau) \rvert}{\sqrt{\langle\lvert\Delta \mathbf{B}(t, \tau)\rvert^2 \rangle}},
\end{equation}
where the temporal increment of the magnetic field $\mathbf{B}$ is defined as $\Delta \mathbf{B}(t, \tau) = \mathbf{B}(t+\tau)-\mathbf{B}(t)$.

\begin{figure}
    \centering
    \includegraphics[width=1\linewidth]{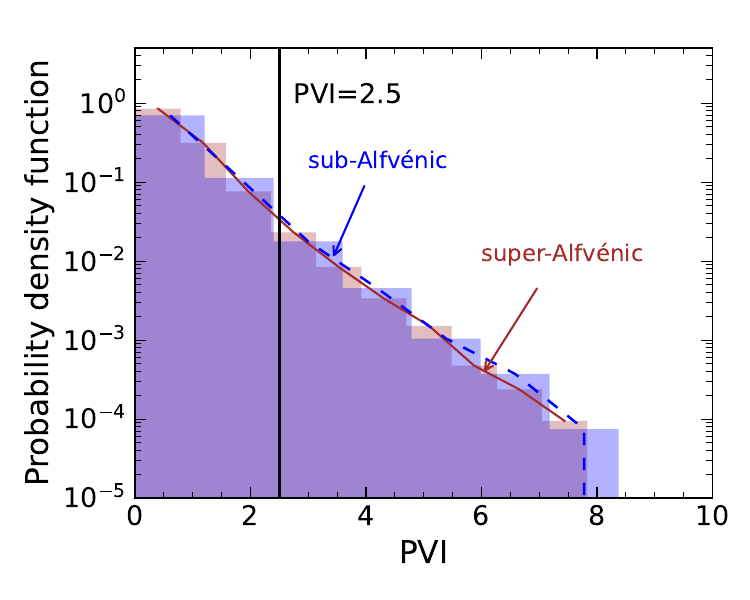}
    \caption{Probability distribution function of PVI in the sub-Alfvénic, and super-Alfvénic solar wind intervals observed by PSP in encounters $8-19$. Values of PVI$>2.5$ represent the non-Gaussian and coherent structures such as current sheets.}
    \label{fig:pvi}
\end{figure}

Here PVI is computed using a moving average of $10$ minutes, denoted by $\langle \dots \rangle$, with a time lag of $\tau=1s$. Fig.~\ref{fig:pvi} shows the distribution of values of PVI for the 
selected samples of sub- and super-Alfv\'enic solar wind. While the turbulence amplitude and variance anisotropy for the sub- and super-Alfv\'enic samples, as
shown above,
have distinctly different features, the PVI distributions 
separated in these two categories 
are very similar and display no distinct trends, consistent to the findings of~\citet{bandyopadhyay2022sub}. The average value of PVI for both the sub- and super-Alfv\'enic cases is $0.75$. Likewise, the most probable values of PVI for both samples are approximately around $0.5$ suggesting the close similarities between these samples with
regard to PVI values. 

\subsection{Magnetic Switchbacks}
Deflections of the magnetic field vector, including the phenomenon of ``switchbacks'', 
may be quantified following~\citet{de2020switchbacks} by the parameter
\begin{equation}
    z=\frac{1}{2}\left(1-\cos \alpha\right),
\end{equation}
where
\begin{equation*}
    \cos \alpha = \frac{\mathbf{B}\cdot \langle \mathbf{B}\rangle}{\lvert \mathbf{B} \rvert \lvert\langle\mathbf{B}\rangle\rvert}
\end{equation*}
and the brackets denote a suitable local or regional average. Here, we again use an averaging interval of 10 minutes. The $z$ variable admits values between $0$ and $1$. Values of $z > 1/2$ indicate that the field is in a polarity-reversed state and lower values correspond to “background” magnetic polarity. In the present analysis, the term ``switchback" refers exclusively to deflections that reverse the polarity i.e., $z > 1/2$.

\begin{figure}
    \centering
    \includegraphics[width=1\linewidth]{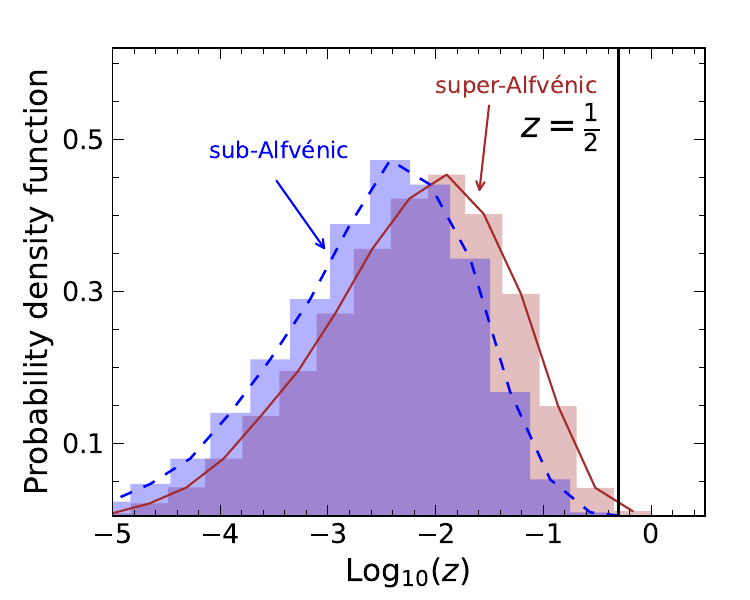}
    \caption{Probability distribution function of switchback parameter ($z$) in the sub-, and super-Alfvénic solar wind intervals observed by PSP in encounters $8-19$. The vertical solid (black) line indicates the value $z = 1/2$, which corresponds to a marginal reversal of polarity of the
    magnetic field. Higher z values represent stronger switchback.}
    \label{fig:switchback_pdf}
\end{figure}

Figure~\ref{fig:switchback_pdf} shows the histograms of the switchback parameter $z$ (in logarithm) for the sub- and super-Alfv\'enic periods. Clearly, the magnetic deflections, as characterized by the $z$ parameter,
are smaller in the sub-Alfvénic periods than in the super-Alfvénic parts, and the deflections in the trans-Alfv\'enic samples lie between these. 
(As found by \citet{bandyopadhyay2022sub}; 
see also \citet{JagarlamudiEA23} for similar results.)
The most probable values of the switchback parameter $\tilde{z}$ for the sub-Alfv\'enic sample is $0.004$, while for the Alfv\'enic and super-Alfv\'enic case, the most probable values $\tilde{z}$  
are $0.008$ and $0.012$ respectively. Similarly, the mean values of the switchback parameter $\overline{z}$ for sub-Alfv\'enic solar wind is $0.012$ with a standard deviation $\sigma_z=0.035$. For the super-Alfv\'enic solar wind  $\overline{z}=0.027$ with $\sigma_z=0.061$.

\begin{table}[t]
\centering
\fontsize{7}{13}\selectfont
\caption{\label{table:average_values} Summary of the average (blue) and most probable (black) values of the several parameters described in the
results section, for the sub-Alfv\'enic ($M_A<0.85$), trans-Alfv\'enic ($0.85\leq M_A \leq 1.15$), and super-Alfv\'enic ($M_A>1.15$) solar wind samples.
}
\begin{tabular}{|c|c|c|c|c|}
\hline
 & & Sub-Alfv\'enic & Alfv\'enic & Super-Alfv\'enic\\
\hline
\centering Turbulence & \textcolor{blue}{$\overline{\frac{\delta B}{B}}$} & \textcolor{blue}{$0.18$} & \textcolor{blue}{$0.22$} & \textcolor{blue}{$0.30$} \\
\cline{2-5}
\centering Amplitude & $\widetilde{\frac{\delta B}{B}}$ & $0.20$ & $0.24$ & $0.38$\\
\hline
\centering Variance & \textcolor{blue}{$\overline{A}_b$} & \textcolor{blue}{$51.09$} & \textcolor{blue}{$31.33$} & \textcolor{blue}{$20.16$}\\
\cline{2-5}
\centering Anisotropy & $\tilde{A}_b$ & $56.23$ & $28.18$ & $13.18$\\
\hline
\centering Partial Variance & \textcolor{blue}{$\overline{\text{PVI}}$} & \textcolor{blue}{0.75} & \textcolor{blue}{0.75} & \textcolor{blue}{0.75}\\
\cline{2-5}
\centering of Increment & $\widetilde{\text{PVI}}$ & 0.62 & 0.48 & 0.40\\
\hline
\centering Switchback & \textcolor{blue}{$\overline{z}$} & \textcolor{blue}{$0.012$} & \textcolor{blue}{$0.015$} & \textcolor{blue}{$0.027$}\\
\cline{2-5}
\centering Parameter & $\widetilde{z}$ & $0.004$ & $0.008$ & $0.012$\\
\hline
\end{tabular}
\end{table}


\section{Discussions and Conclusions} \label{sec:discussion}

As PSP orbits descended more deeply
into the solar atmosphere, the occurrence rate
of sub-Alfvenic intervals has sharply increased, being 
less than 10\% at 30-40 $R_s$ and increasing to $>$70\% at 14 $R_s$ \citep{chhiber2024alfven}.
Indeed, 
in more recent later 
orbits 
PSP has sampled extended periods of sub-Alfv\'enic solar wind.
Comparing the sub-Alfv\'enic periods with the trans-Alfv\'enic and super-Alfv\'enic periods (see Table~\ref{table:average_values}), we find that in the sub-Alfv\'enic solar wind:
(i) the normalized 
turbulence amplitude decreases;
(ii) anisotropy increases;
(iii) intermittency as measured by PVI is almost unchanged; and
(iv)
the angular deflections of the magnetic field are on average weakening, indicating the lack of switchbacks~\citep{ruffolo2020shear}. 
Overall the 
behavior is consistent with the 
initial study by \citet{bandyopadhyay2022sub}
except that in the present study we have
much better statistical coverage of the sub-Alfv\'enic samples.

Once again we can see that the sub-Alfv\'enic
wind more closely resembles properties that one might associate with coronal conditions.
Often the corona is modeled based on 
properties expected for low beta, highly anisotropic  plasma , which may be dominated by incompressive fluctuations such as Alfv\'enic 
fluctuations (or waves). A typical coronal model of this type is based on Reduced Magnetohydrodynamics (RMHD) as employed in, e.g., 
\citet{EinaudiVelli99,GomezEA00-corona,OughtonEA01}. 
In contrast, the super-Alfv\'enic solar wind is much less anisotropic and is typically modeled
in the fluid regime by compressible MHD models without the reduction in dimensionality leading to RMHD.
In this regard it has been shown that the 
validity of the RMHD approximation is particularly sensitive to the presence of component variance in the direction parallel to the regional mean magnetic field \citep{DmitrukEA05-rmhd}.
From this perspective it appears that, as it approaches the sun, 
PSP is sampling not only a complex 
transition from super-Alfv\'enic to sub-Alfv\'enic wind \citep{chhiber2024alfven}, but also 
a transition from a large plasma beta, more fully three dimensional compressible MHD plasma (at scales much larger than the ion inertial scale) to 
a plasma better described as highly anisotropic, less compressible and described by Reduced MHD.
We note that the 
anisotropy measure shown in 
Fig.~\ref{fig:Anisotropy_pdf}
is essentially the reciprocal of the quantity sometimes called magnetic compressibility~\citep{kiyani2012enhanced}; it's behavior is consistent with greater compressibility in the super-Alfv\'enic regime. 
Other signatures of this transition, such as the
presence of co-rotation as a signature of magnetic dominance in  the corona, are anticipated
\citep{weber1967angular,kasper2019alfvenic,chhiber2025effect}.

\begin{acknowledgments}
Acknowledgements: This work is supported at the University of Delaware in part by the PSP/ISOIS 
project
through subcontract SUB0000165 from Princeton to the University of Delaware, and by the PUNCH project under subcontract N99054DS. 
JG is supported by the Delaware
NASA Space Grant program grant number
80NSSC20M0045 at the University of Delaware. AU, RC and SA are partially 
supported in part by 
NASA under grant number 
80NSSC22K1639. RC is also supported by NASA grant number 80NSSC22K1020. This research is also supported in Thailand by the National Science and Technology Development Agency (NSTDA) and the National Research Council of Thailand (NRCT) through the High-Potential Research Team Grant Program (N42A650868).
\end{acknowledgments}

%





\section{Appendix}
\begin{figure*}
    \centering
    \includegraphics[width=1.\linewidth]{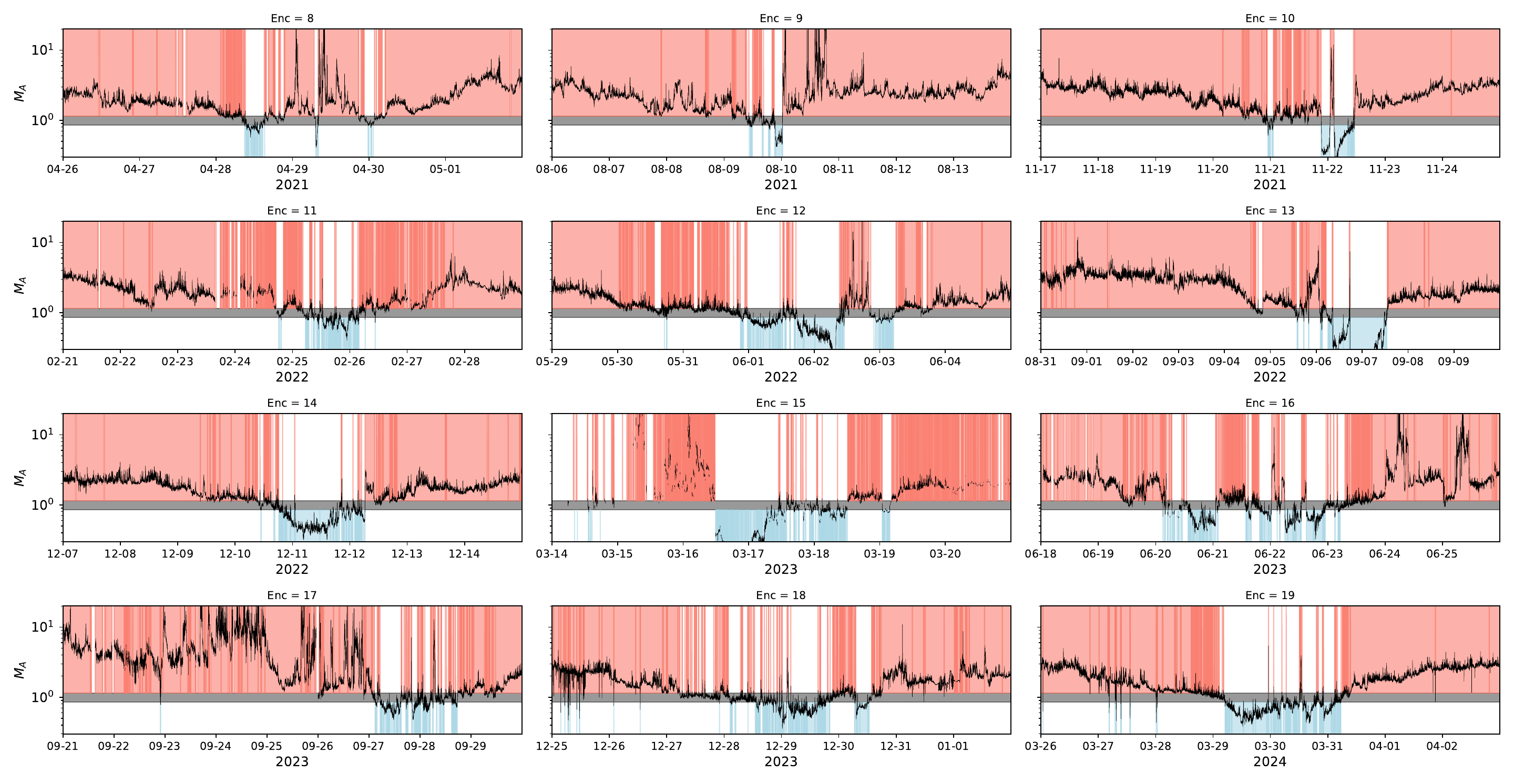}
    \caption{An overview of the Alfvén Mach number $M_A = V_R/V_A$ where $V_A$ is the Alfv\'en speed, as observed by the Parker Solar Probe (PSP) for encounters $8-19$. Sub- and super-Alfv\'enic intervals are shown by the shaded regions colored light blue and light red respectively. In addition, shaded gray region extending $0.85 \leq M_A \leq 1.15$ represents what we have termed 
    the ``trans-Alfv\'enic dataset'' 
    isolated to contrast the behavior between sub- and super-Alfv\'enic solar wind.}
    \label{fig:PSP_data}
\end{figure*}

Here we provide an overview of the PSP datasets from encounters $8$--$19$. In Fig.~\ref{fig:PSP_data}, the datasets are shown characterized based on Alf\'ven Mach number. Sub- and super-Alv\'enic data samples are represented by the shaded regions with lighter shades of blue and red respectively, with an additional gray region separating the trans-Alfv\'enic dataset ($0.85\leq M_A \leq 1.15$). The dates of the intervals from each enouncter used in this study are given in Table \ref{tab:tableI}.

\begin{table}
\caption{\label{tab:tableI}Start and end times of the $4$ days of dataset (centered at the perihelion) analyzed for Parker Solar Probe (PSP) encounters $8$--$19$.}
\begin{ruledtabular}
\begin{tabular}{||c | c | c||} 
 \hline
 Enc & Start (UTC) & End (UTC) \\ [0.5ex] 
 \hline\hline
 8 & 2021-04-27 06:00:30 & 2021-05-01 06:00:30 \\ 
 \hline
 9 & 2021-08-07 18:00:30 & 2021-08-11 18:00:30 \\
 \hline
 10 & 2021-11-19 09:00:30 & 2021-11-23 09:00:30 \\
 \hline
 11 & 2022-02-23 12:00:30 & 2022-02-27 12:00:30 \\
 \hline
 12 & 2022-05-31 00:00:30 & 2022-06-04 00:00:30 \\
 \hline
 13 & 2022-09-04 06:00:30 & 2022-09-08 06:00:30 \\ 
 \hline
 14 & 2022-12-09 12:00:30 & 2022-12-13 12:00:30 \\
 \hline
 15 & 2023-03-15 21:00:30 & 2023-03-19 21:00:30 \\
 \hline
 16 & 2023-06-20 00:00:30 & 2023-06-24 00:00:30 \\
 \hline
 17 & 2023-09-26 00:00:30  & 2023-09-30 00:00:30 \\
 \hline
  18 & 2023-12-27 00:00:30 & 2023-12-31 00:00:30 \\
 \hline
 19 & 2024-03-28 00:00:30 & 2024-04-01 00:00:30 \\
 [1ex] 
 \hline
\end{tabular}
\end{ruledtabular}
\end{table}

\bibliography{sample631}{}
\bibliographystyle{aasjournal}



\end{document}